# Critical issues in the formation of quantum computer test structures by ion implantation

T. Schenkel[a], C. C. Lo[b], C. D. Weis[a], A. Schuh[a], A. Persaud[a], and J. Bokor[b]

[a]Accelerator and Fusion Research Division, Lawrence Berkeley National Laboratory,
Berkeley, CA 94720, USA

[b]Department of Electrical Engineering and Computer Science, University of California,
Berkeley, CA 94720, USA

The formation of quantum computer test structures in silicon by ion implantation enables the characterization of spin readout mechanisms with ensembles of dopant atoms and the development of single atom devices. We briefly review recent results in the characterization of spin dependent transport and single ion doping and then discuss the diffusion and segregation behaviour of phosphorus, antimony and bismuth ions from low fluence, low energy implantations as characterized through depth profiling by secondary ion mass spectrometry (SIMS). Both phosphorus and bismuth are found to segregate to the $SiO_2/Si$ interface during activation anneals, while antimony diffusion is found to be minimal. An effect of the ion charge state on the range of antimony ions, $^{121}Sb^{25+}$, in $SiO_2/Si$ is also discussed.

## I. Introduction

Quantum computing promises to revolutionize information technology, and the viability of a large array of potential implementation proposals is currently being tested. Spins (electron and nuclear) of donor atoms in silicon are attractive quantum bit (qubit) candidates due to long coherence times and the finesse of silicon nanotechnology [1]. To date, reliable single spin readout and control has not been achieved with donors in silicon. Yet, significant progress has been made towards this crucial demonstration [2-4]. We recently reported the observation of spin dependent scattering in $^{28}Si$ based accumulation channel field effect transistors (aFets) through measurements of electrical detection of magnetic resonance (EDMR). Here, the channels of micron scale transistors had been implanted with $^{121}Sb$ ions [2]. The observed resonant current changes dI/I were a few times $10^{-7}$ at 5 K, about a factor of 100 smaller then claimed in studies with bulk doped devices. The latter might have been actually dominated by microwave heating effects, while in our results the scaling of the resonant current changes with gate bias [2] and temperature, dI/I~1/T [5], clearly point to neutral impurity scattering in the channel of implanted devices. The temperature dependence is due to the increased polarization of donor atoms at lower temperature, and the linear dependency of the resonant current change on donor polarization.

Theoretical analysis of the scattering problem has led to the proposal of a technique for single donor electron spin readout that is both projective and quantum non-demolition within the nuclear spin relaxation time, $T_{1n}$ [6]. Here, the quantum information in the electron spin is transferred to the nuclear spin through a series of RF and microwave pulses. The hyperfine splitting of the electron spin resonance lines effectively map the spin state of the nucleus to the magnetic field in the EDMR setup where a resonant current change can be observed in EDMR. For single donors, the detection of a single resonant line, rather then the full spectrum of hyperfine split lines that is observed in ensembles of donors then constitutes a readout of the nuclear spin state of the single donor. For this to work, it is required that the nuclear spin must not flip during the EDMR measurement. Estimated measurement times under realistically achievable conditions are >1 ms. While the donor electron spin state mixes quickly when a current of conduction electrons scatters off the donor, the nuclear spin is rather isolated, and it was recently found that $T_{1n}$ does not increase when the donor electron spin is continuously flipped by a resonant microwave field [5].

Moreover, the state transfer of the spin information from the electron spin to the nuclear spin and then back to the electron spin was recently demonstrated with ensembles of $^{31}$P atoms in a $^{28}$Si crystal [7]. Here, the nuclear spin coherence time could be extended to $T_{2n}>1.75$ s at 5º K using refocusing pulse sequences. This remarkably long nuclear spin coherence time underscores the potential of donor spin qubits in silicon.

In order to harvest this fundamental property for quantum information processing, arrays of single donor atoms have to be integrated with high precision into low noise readout transistors. Towards this end, we recently reported a technique for reliable single ion impact sensing in readout transistors. In tests with micron scale transistors operated at room temperature, we showed that single ion hits lead to distinct steps in the source – drain currents of transistors [8]. The ion beam can now be aligned to an opening in the gate stack of the transistor with an in situ scanning force microscope (SFM). The aperture in the cantilever of the SFM functions as a dynamic shadow mask. This way, arrays of single dopant atoms can be implanted into devices. Alternatively, the response of transistors to incident ions can also be mapped with high spatial resolution [9].

Based on these developments with micron scale transistors, we now formed a first generation of 100 nm scale transistors in a FinFet geometry (Fig. 1) in SOI (silicon on insulator).

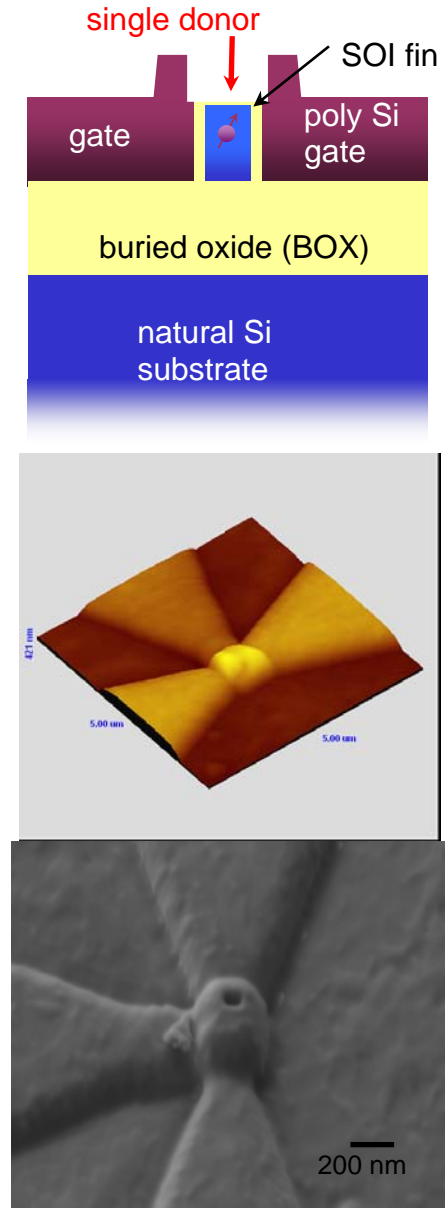

Figure 1: Schematic of the FinFet with hole in the gate (a, top), and in situ SFM (b, mid) and ex situ SEM images (bottom) of a FinFet after gate surgery in a dual beam FIB (c, bottom).

The device silicon consists of a 150 nm thick epi-layer of $^{28}$Si on a 100 nm natural silicon layer. Figures 1 b) and c) show an in situ SFM image taken in the implant chamber and an ex situ electron micrograph of a FinFet after "gate surgery", respectively. In the latter process, a hole is opened in the gate stack using a combination of ion beam drilling and electron beam assisted etching in a dual beam FIB instrument [8]. Tungsten metallization of transistors allows repeated cycles of ion implantation, annealing for damage repair and dopant activation and transport spectroscopy.

One critical choice in test device development concerns the dopant species. Kane used phosphorus in his original proposal [1], and he specified a dopant spacing of 20 nm. In a coupled qubit device, single dopant atoms have to be placed into desired locations with spacings of 20 nm and minimal placement uncertainty. Range straggling as well as diffusion during post-implantation annealing contribute to the error budget in dopant placement. Besides phosphorus, arsenic, antimony and bismuth are also shallow, substitutional donors in silicon, and the weakly bound electrons of either of these dopants could be used to encode quantum information. We now describe results from a process metrology study using SIMS depth profiling for the characterization of dopant segregation and diffusion during standard rapid thermal annealing (RTA) and report a charge state effect on the stopping of multiply charged Sb ions in $SiO_2$/Si.

**II. Experimental setup**
Samples were highly resistive silicon wafers (>1 kOhm cm, 100), covered with a 10 or 20 nm thick oxide layer from a standard dry oxidation process. For phosphorus implantations, we used $^{28}$Si epi layers (>99.9 $^{28}$Si) to avoid the $^{30}$SiH mass interference with $^{31}$P, which might have skewed previous studies of low dose $^{31}$P depth profiles in natural silicon. One $^{31}$P implant series was conducted with $^{28}$Si samples covered with silicon nitride layers instead of $SiO_2$. Antimony and bismuth were implanted at 120 and 60 keV, respectively and with fluences of $2 \times 10^{11}$ cm$^{-2}$ and $4 \times 10^{11}$ cm$^{-2}$ under zero degree tilt angle. Post implant annealing was performed in an AGA Heatpulse rapid thermal annealing (RTA) chamber with temperatures of 1000º C for 10 s in $N_2$ ambient.

Highly charged ions were extracted from the Electron Beam Ion Trap (EBIT) at Lawrence Berkeley National Laboratory, and reached the target chamber after momentum analysis in a 90° bending magnet [10]. Singly charged ions were implanted with a commercial implanter. SIMS analysis was performed by Evans Analytical.

**III. Results and discussion**
In Fig. 2, we show SIMS depth profiles for low dose 60 keV $^{31}$P$^+$ implants into $^{28}$Si samples covered with a 10 nm silicon nitride layer (a) or a 10 nm $SiO_2$ layer (b). SIMS analysis of low dose phosphorus implants is challenging. Peak concentrations are only $5 \cdot 10^{16}$ cm$^{-3}$, and the SIMS profiles are noisy. The as implanted profile peaks at a depth of about 60 nm below the surface. Dopants diffuse and segregate during RTA through coupling to specific defects. Phosphorus is an interstitial diffuser [11-14]. Interstitials are injected from the $SiO_2$/Si interface into silicon during annealing under mildly oxidizing conditions (e. g. from residual water in the RTA chamber), while vacancies are injected from silicon nitride. The former enhance, while the latter retard $^{31}$P diffusion. Phosphorus atoms segregate to the SiO$_2$/Si interface, where they are bond and electrically inactive [13]. It is unclear to date if the segregation enhancement for low concentrations of phosphorus is an intrinsic property of the SiO$_2$/Si interface, or if it can be eliminated for "perfect" interfaces perfectly non-oxidizing conditions [14]. Significant range straggling is also apparent in the depth profiles. Evidently, the symmetrical scattering condition in the close mass match of silicon and phosphorus leads to the largest range straggling of all shallow group V donors in silicon. Bismuth is the heaviest shallow donor in silicon and the SIMS spectra in Figure 3 show remarkably narrow widths. However, and much to our surprise, Bi atoms were also found to segregate to the SiO$_2$/Si interface during RTA. The mechanism for this is unclear. Bi is a vacancy diffuser, and we speculate that the intense radiation damage from individual collision cascades results in such massive damage that the resulting defect structure aids Bi segregation to the interface. In Fig. 4 we show depth profiles for Sb implants. We previously found that RTA does not lead to any significant dopant movement for Sb [15], making Sb an optimal choice for single atom device development. Arsenic was not tested here, since we use As implants for the formation of highly n-type doped source – drain contacts in the readout transistors and want to avoid potential ambiguities between these and the qubit implants.

Finally, we implanted $^{121}$Sb$^{25+}$ ions and found that their range at an implant energy of 120 keV was significantly reduced compared to $^{121}$Sb$^{1+}$ implants (Figure 4), from a peak depth of 50 nm to 30 nm. A fluence dependence of the profiles in Figure 5 shows no effect of the peak depth on the implanted fluence, excluding accumulation of damage in the oxide or charge state enhanced sputtering as reasons for the reduced range. The latter would be expected to lead to an increased depth in any case. Also, the fluence dependence shows that the strong surface peak is an artefact from the SIMS measurement, maybe from a mass interference from an oxide molecular ion like (SiO$_2$)$_2$H$^-$.

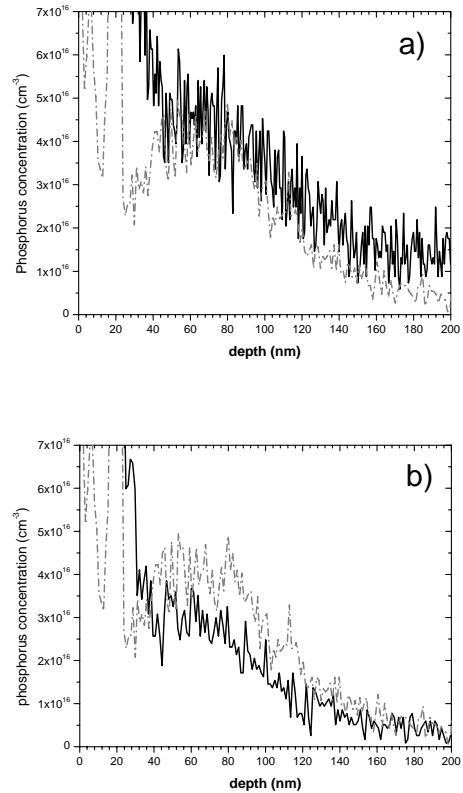

Figure 2: SIMS depth profiles of $^{31}$P atoms implanted into $^{28}$Si samples (implant energy 60 keV). a) As implanted profile (dash-dot) (fluence=4x10$^{11}$ cm$^{-2}$) and profile after annealing in the presence of an Si$_3$N$_4$ (solid line). The fluence for the implant with nitride layer was 2x10$^{11}$ cm$^{-2}$ so we multiplied the concentration in SIMS profile by two. b) As implanted $^{31}$P (as in a)) and profile after annealing in the presence of an SiO$_2$ layer (solid line).

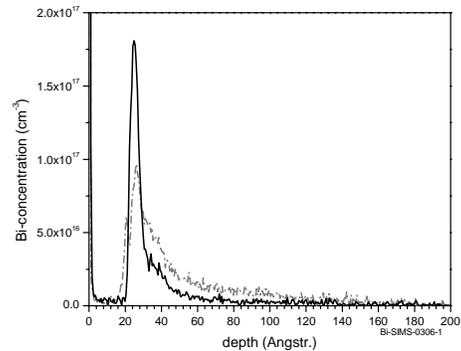

Figure 3: SIMS depth profiles for Bi implants (2x10$^{11}$ cm$^{-2}$, 60 keV) as implanted (dashed line) and after annealing in the presence of an SiO$_2$ layer (solid).

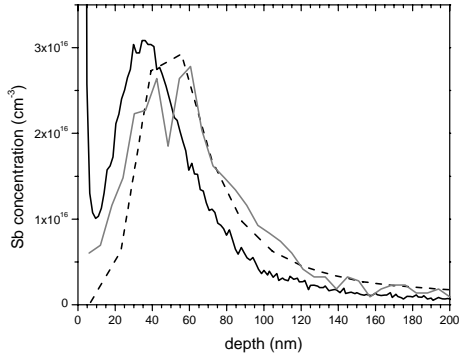

Figure 4: SIMS depth profiles of $^{121}$Sb$^{1+}$ (fluence = $2\times10^{11}$ cm$^{-2}$) (grey, solid) after annealing and a simulated profile of as implanted $^{121}$Sb (dashed) [14], together with a profile of as implanted $^{121}$Sb$^{25+}$ (fluence=$7.1\times10^{11}$ cm$^{-2}$, profile scaled for comparison with Sb$^{1+}$). The implant energy was 120 keV.

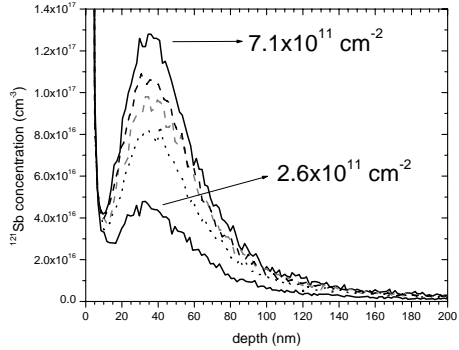

Figure 5: SIMS profiles of $^{121}$Sb$^{25+}$ (120 keV) as a function of fluence.

We attribute this range reduction to charge state dependent stopping of the multiply-charged Sb ions in the oxide layer. Charge state enhanced stopping of relatively slow ions in insulators has been observed before for Ar$^{16+}$ in thin CaF$_2$ films [16] and for a series of high charge state ions in thin films of amorphous carbon [17]. The energy loss increase can be attributed to the finite relaxation time (of order 10 fs) of the transient hollow atom inside the solid and reduced screening conditions that lead to increased momentum transfer to target electrons and nuclei during the relaxation time. Systematic studies of this effect as a function of oxide thickness and ion charge state are in progress. If indeed reduced screening during a transient relaxation time is responsible for the enhanced stopping, then it might be the case that the on average larger impact parameters under these conditions lead to reduced straggling for implantation of high charge state, low energy ions in dielectric materials such as thin gate oxide layers.

For sensitive quantum transport devices, damage to the oxide layer from the (single) ion implantation step is a critical issue. It can be expected that this damage is more severe for heavier and more highly charged dopant ions compared to $^{31}$P$^{1+}$, posing a complementary advantage for the fast diffusing $^{31}$P over e. g. Sb. The effectiveness of damage repair in the oxide under optimized annealing or re-oxidation conditions is subject of ongoing studies.

### V. Conclusions
The development of quantum computer test structures in silicon is progressing with studies of ever smaller ensembles of dopant atoms and experiments with 100 nm scale devices that are compatible with iterative single atom doping. Basic ion solid interactions play a crucial role in the selection of optimal conditions for qubit donor selection and placement. We find that antimony is an optimal choice for high precision single donor placement due to minimal diffusion and segregation during annealing and small range straggling, but questions regarding more severe implant damage and the effectiveness of damage repair during annealing have to be resolved. In contrast both phosphorus and bismuth are found to exhibit significantly enhanced dopant movement during annealing. For phosphorus, defect engineering, i. e. selective acting on the vacancy – interstitial balance and supersaturation of an interface near layer with vacancies, e. g. by carbon co-implantation or MeV Si-ion implantation are possible routes to suppress dopant segregation to the $SiO_2$/Si interface that is driven by interstitials. Charge state enhanced stopping for highly charged Sb ions in $SiO_2$ might enable implantation with reduced range straggling due to a shift to larger impact parameters for collisions during charge state relaxation, and ongoing studies aim at quantifying this effect.

**Acknowledgements:**
We thank the staff of the Molecular Foundry and the National Center for Electron Microscopy at LBNL for their support. This work was in part supported by NSA under Contract No. MOD 713106A, the National Science Foundation through NIRT Grant No. CCF-0404208, and by the Director, Office of Science, of the Department of Energy under Contract No. DE-AC02-05CH11231.